\documentclass[twoside]{article}
\usepackage[utf8]{inputenc}
\usepackage{enumitem}
\usepackage{hyperref}
\usepackage{geometry}
\usepackage{graphicx}
\usepackage{float}
\usepackage{changepage}
\usepackage{amsmath}
\usepackage{multirow}
\usepackage{multicol}
\usepackage{makecell}
\usepackage{tabularx}
\usepackage{array}
\usepackage{caption}
\usepackage{needspace}

\usepackage[numbers]{natbib}

\title{DeepMultiConnectome: Deep Multi-Task Prediction of Structural Connectomes Directly from Diffusion MRI Tractography}

\author{
  Marcus J. Vroemen\textsuperscript{1,2}\thanks{Corresponding author: m.j.vroemen@student.tue.nl}, 
  Yuqian Chen\textsuperscript{2,3}, 
  Yui Lo\textsuperscript{2,3,4}, 
  Tengfei Xue\textsuperscript{4}, \\
  Weidong Cai\textsuperscript{4}, 
  Fan Zhang\textsuperscript{5}, 
  Josien P.W. Pluim\textsuperscript{1}, 
  Lauren J. O'Donnell\textsuperscript{2,3,6}
}

\date{}  

\usepackage{fancyhdr}
\begin{document}

\maketitle

\vspace{-1.5em}  

{\centering
\small 
\textsuperscript{1} Eindhoven University of Technology, Eindhoven, The Netherlands \\
\textsuperscript{2} Brigham and Women's Hospital, Boston, USA \\
\textsuperscript{3} Harvard Medical School, Boston, USA \\
\textsuperscript{4} The University of Sydney, Sydney, Australia \\
\textsuperscript{5} University of Electronic Science and Technology of China, Chengdu, China \\
\textsuperscript{6} Harvard-MIT Health Sciences and Technology, Cambridge, USA \par
\vspace{0.5em}
}

\vspace{1em}
\begin{adjustwidth}{3em}{3em}
{\normalsize\textbf{Abstract.} Diffusion MRI (dMRI) tractography enables in vivo mapping of brain structural connections, but traditional connectome generation is time-consuming and requires gray matter parcellation, posing challenges for large-scale studies. We introduce DeepMultiConnectome, a deep-learning model that predicts structural connectomes directly from tractography, bypassing the need for gray matter parcellation while supporting multiple parcellation schemes. Using a point-cloud-based neural network with multi-task learning, the model classifies streamlines according to their connected regions across two parcellation schemes, sharing a learned representation. We train and validate DeepMultiConnectome on tractography from the Human Connectome Project Young Adult dataset ($n=1000$), labeled with an 84 and 164 region gray matter parcellation scheme. DeepMultiConnectome predicts multiple structural connectomes from a whole-brain tractogram containing 3 million streamlines in approximately 40 seconds. DeepMultiConnectome is evaluated by comparing predicted connectomes with traditional connectomes generated using the conventional method of labeling streamlines using a gray matter parcellation. The predicted connectomes are highly correlated with traditionally generated connectomes (\textit{r} = 0.992 for an 84-region scheme; \textit{r} = 0.986 for a 164-region scheme) and largely preserve network properties. A test-retest analysis of DeepMultiConnectome demonstrates reproducibility comparable to traditionally generated connectomes. The predicted connectomes perform similarly to traditionally generated connectomes in predicting age and cognitive function. Overall, DeepMultiConnectome provides a scalable, fast model for generating subject-specific connectomes across multiple parcellation schemes.}

\vspace{1em}
\end{adjustwidth}

\section{Introduction} \label{sec:1}

The human connectome, a comprehensive map of neural connections in the brain, is fundamental to understanding brain organization, function, and dysfunction \cite{Sporns2005, Toga2012}. Diffusion magnetic resonance imaging (dMRI) tractography is currently the only non-invasive imaging modality that enables in vivo mapping of structural connections. By modeling white matter streamlines with tractography, the connectivity between predefined gray matter regions of interest (ROIs) can be estimated. This connectivity is often summarized in a connectivity matrix, or “structural connectome,” that characterizes the brain’s structural connections. Analysis of these connectomes has provided valuable insights into cognitive processes, neurological disorders, and brain development \cite{Bassett2009, Cao2016, Gruber2023, Wainberg2024}. Beyond their descriptive utility, structural connectomes are increasingly employed to make predictions about individual differences in cognition and behavior \cite{Bernstein-Eliav2024, Shen2017}. Structural connectomes are also useful for predicting brain age, an overall marker of brain health \cite{Baecker2021, Karimi2024}.

\begin{figure}[h]
    \centering
    \includegraphics[width=0.7\linewidth]{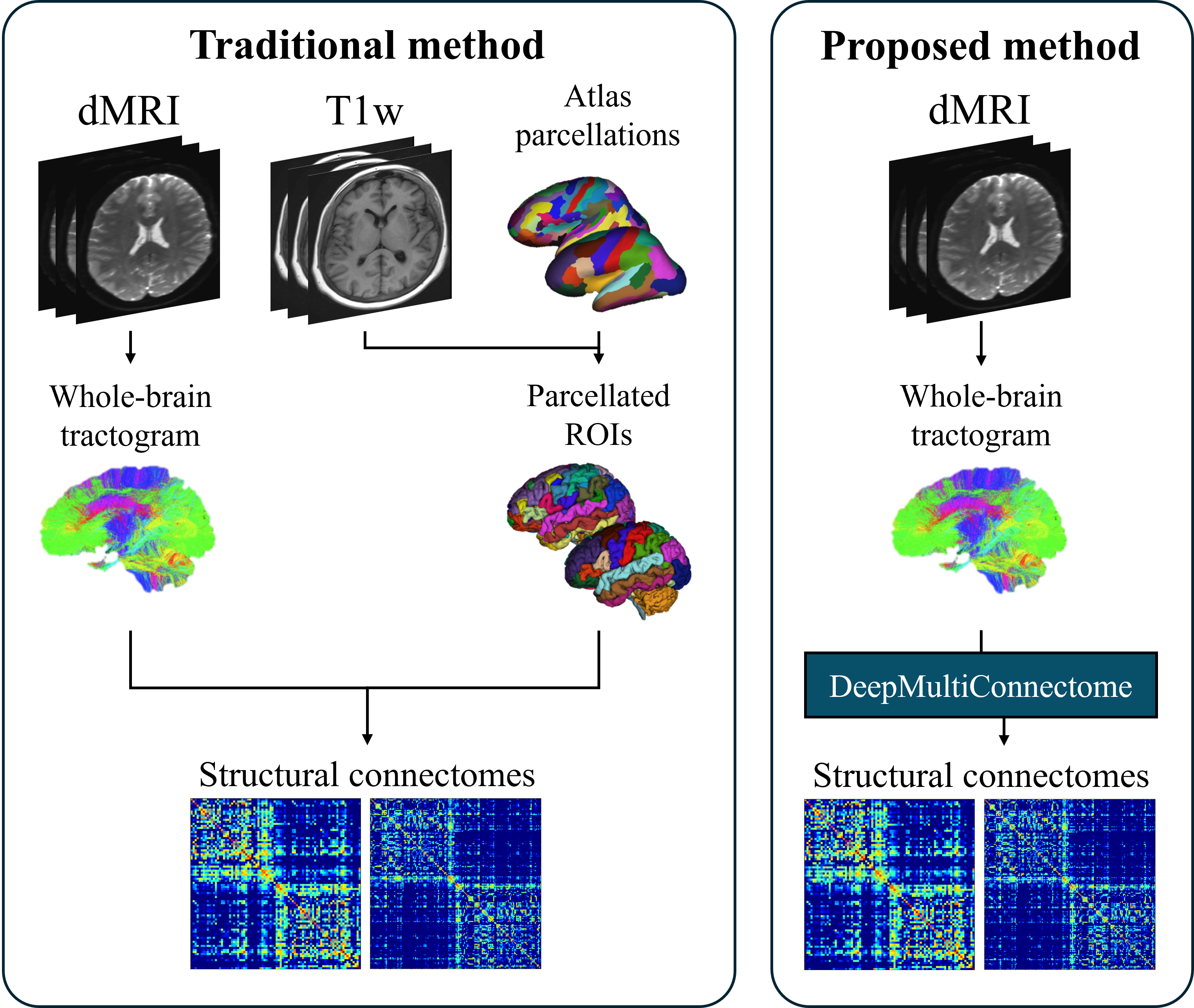}
    \caption{Structural connectome construction using the traditional and proposed method. The traditional method requires an anatomical parcellation of the brain, which can be computationally expensive. Our method avoids this by using a deep learning network to predict the anatomical regions that streamlines connect.}
    \label{fig:1}
\end{figure}

However, generating structural connectomes is a computationally intensive process that requires both the computation of whole-brain tractograms and the parcellation of cortical and subcortical gray matter based on anatomical atlases (Figure~\ref{fig:1}). Traditional cortical parcellation methods, such as FreeSurfer, can take between 5 to 24 hours per subject, depending on the computational resources and data quality \cite{Fischl2012}. This processing time can create a critical bottleneck for large-scale neuroimaging studies that involve thousands of participants. While deep learning models like FastSurfer have reduced this parcellation time to minutes, the output is based on a single parcellation atlas \cite{Henschel2020}, limiting flexibility across diverse research needs \cite{Arslan2018}. A method that bypasses gray matter parcellation and predicts structural connectomes directly from tractography data could accelerate processing, offering substantial advantages in large dataset handling. Although prior methods have been developed to predict structural connectomes from raw diffusion MRI data or from non-imaging phenotype data \cite{Liu2025, Sarwar2020}, these existing approaches do not leverage tractography data as input and do not predict multiple connectomes. In this work, our overall goal is to predict multiple structural connectomes directly from tractography streamline data in a computationally efficient manner.

Advances in deep learning have demonstrated the potential for fast and consistent segmentation of the brain’s white matter structural connections into anatomically defined fiber tracts \cite{Wasserthal2018, Xue2023-b, Zhang2020}. Point-cloud-based networks, which represent tractography streamlines as ordered points in 3D space, have further improved performance by naturally preserving streamline geometry, ensuring invariance to point ordering, and enhancing computational efficiency and robustness across diverse populations \cite{Astolfi2020, Chen2023-a, Chen2022, Xue2023-a, Xue2023-b}. However, the application of such networks for classifying tractography streamlines based on gray matter regions to predict structural connectomes remains unexplored.

To generate a structural connectome, information is needed on which ROIs each streamline connects to. By predicting the ROI pair a streamline connects, it is possible to bypass the computationally intensive gray matter parcellation step. Building on previous deep-learning tractography classification models, we propose adopting this strategy to classify streamlines by their connected ROIs rather than anatomical tracts. In this approach, the ROIs associated with each streamline are used as training labels, corresponding to the desired parcellation scheme.

We propose to realize further efficiencies through a multi-task learning approach, extending point-cloud-based classification to simultaneously predict connectomes across different gray matter parcellation schemes. Multi-task learning enables a single model to perform multiple related tasks at once, allowing shared representations to improve generalization and efficiency, as described in a recent review \cite{Zhang2018-b}. This approach has been successfully applied in various areas, including computer vision, natural language processing, and medical imaging, where leveraging shared knowledge across tasks improves performance \cite{Karimi2024, Lo2024-a, Zhang2018-b}. Given the inherent similarity between predicting connectomes for different parcellation schemes---leveraging the same underlying tractography streamline data but with varying ROI definitions---this approach reduces computational cost while potentially enhancing prediction accuracy using shared information across parcellation schemes.

This study proposes a deep-learning model capable of predicting structural connectomes directly from tractograms without requiring gray matter parcellation. By training a point cloud classification model to assign streamlines to gray matter regions, our approach enables the fast prediction of connectomes across multiple parcellation schemes. This model has the potential to accelerate large-scale neuroimaging studies, facilitating flexible, efficient analyses across various parcellation schemes.

\section{Methods} \label{sec:2}
In this section, we outline the methodology used in this study. We first describe the dataset~\ref{sec:2.1}, including the preprocessing steps to obtain tractography and structural connectomes~\ref{sec:2.2}. Next, we provide details on the proposed connectome prediction model~\ref{sec:2.3} and its implementation~\ref{sec:2.4}.

\subsection{Dataset} \label{sec:2.1}

We used the Human Connectome Project Young Adult (HCP-YA) dataset data to generate whole-brain tractograms and structural connectomes. The HCP-YA study collected high-quality neuroimaging and behavioral data from healthy adults aged 22-35 \cite{VanEssen2013}. We employed diffusion and structural MRI data from 1,000 subjects, plus an additional 45 subjects scanned twice as part of a test-retest set. The dMRI data was acquired on a customized Siemens 3T ``Connectome Skyra'' at Washington University using a spin-echo echo-planar imaging (EPI) sequence. Acquisition parameters included: TR = 5520 ms, TE = 89.5 ms, FA = 78°, 1.25 mm isotropic voxel size, and FOV = 210 × 180 mm\(^2\). A total of 270 diffusion-weighted images were acquired, evenly distributed at three shells with b = 1000/2000/3000 s/mm\(^2\) \cite{Sotiropoulos2013}.

Preprocessing of the dMRI data included correction for eddy currents, head motion, and EPI distortions, followed by co-registration with T1-weighted (T1w) structural data \cite{Glasser2013}. The T1w images were acquired at TR = 2400 ms, TE = 2.14 ms, and 0.7 mm isotropic voxel size. These structural images were registered to MNI152 space using nonlinear FSL FNIRT transformations \cite{VanEssen2013}, which were later used to register the whole-brain tractograms to MNI space.

The HCP-YA dataset provided two FreeSurfer parcellations for all subjects \cite{Fischl2012}. They contain either 34 cortical regions per hemisphere, based on the Desikan-Killiany atlas, or 74 cortical regions per hemisphere, based on the Destrieux atlas \cite{Desikan2006, Destrieux2010}. Additionally, both contain a segmentation of the cerebellum cortex and 7 subcortical regions (the thalamus (proper), caudate, putamen, pallidum, hippocampus, amygdala, and nucleus accumbens) for each hemisphere. This results in one parcellation scheme with 84 ROIs and a second with 164 ROIs.

In addition, we used cognitive performance measures from the NIH Toolbox, including the NIH Toolbox Picture Vocabulary Test (TPVT), the Oral Reading Recognition Test (TORRT), the Flanker Attention and Inhibitory Control Test (TFAT), the Cognitive Shift Test (TCST), the List Sorting Working Memory Test (TLST), and the Pattern Comparison Processing Speed Test (TPIT) \cite{Gershon2013, Weintraub2013}.

\subsection{Tractography and connectivity matrix generation} \label{sec:2.2}
\begin{figure}[h]
    \centering
    \includegraphics[width=0.8\linewidth]{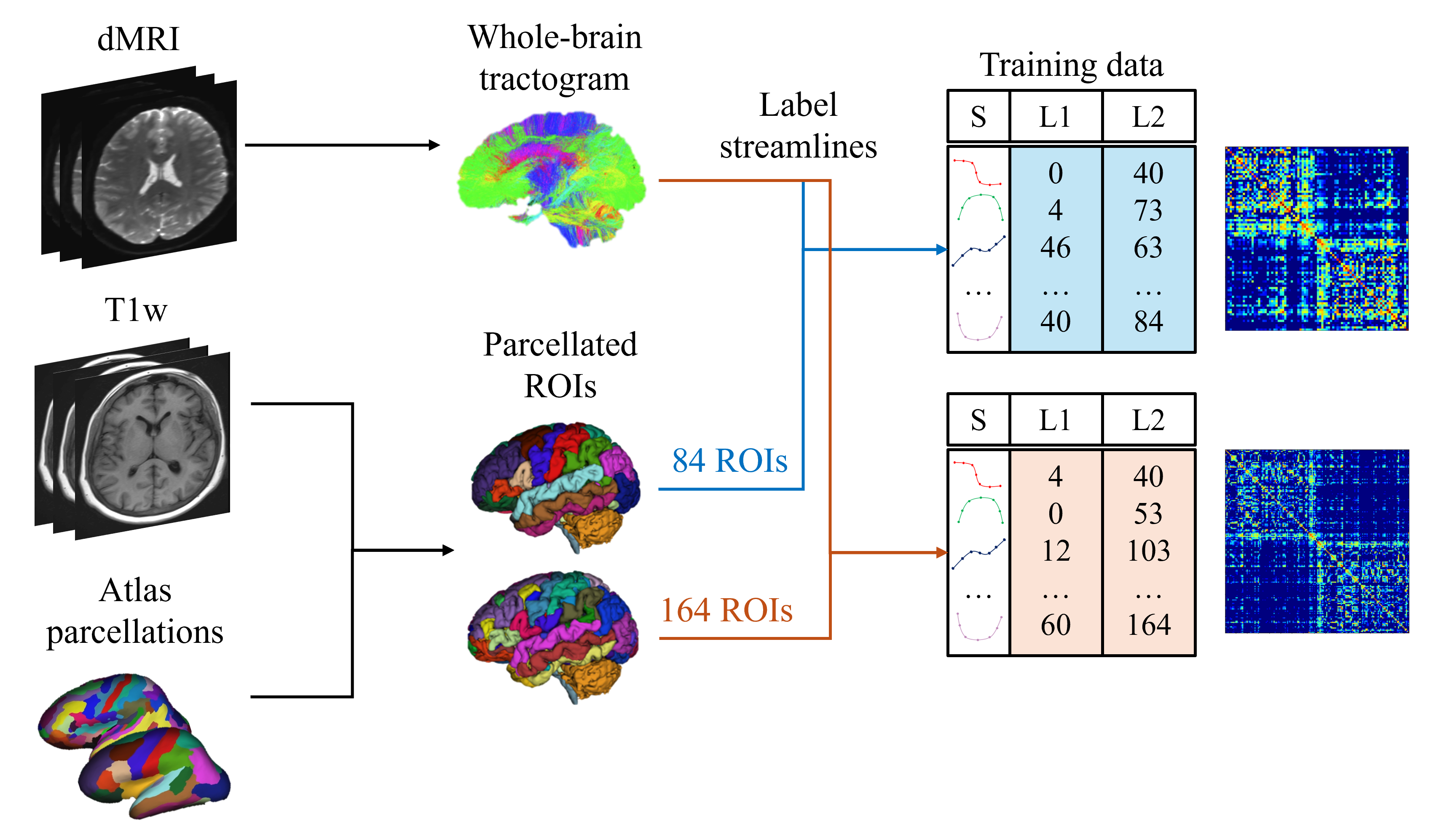}
    \caption{Diagram of the data preparation pipeline, showing how the MRI data is used to generate whole-brain tractograms and structural connectomes. The two different gray matter parcellations are used to give two label sets to each streamline and generate two structural connectomes. The streamlines (S) and assigned labels (L1 and L2) form the training dataset.}
    \label{fig:2}
\end{figure}

Standard processing methods were applied to create tractography and ground truth connectome matrices using two gray matter parcellations (Figure~\ref{fig:2}). This labeled tractography data served as the training data for the connectome prediction model.

\subsubsection{Tractography} \label{sec:2.2.1}
Whole-brain tractograms were generated using probabilistic tractography with MRtrix3 \cite{Tournier2019}. Following standard procedures, the T1w image was used to create a gray matter-white matter (GMWM) interface for seeding, while dMRI data were used to estimate fiber orientation distributions (FODs) before streamline generation. The pipeline starts with deriving a tissue-type segmentation from the T1w image, which is used for Anatomically-Constrained Tractography (ACT) \cite{Smith2012} and to construct the GMWM interface mask. The interface was then co-registered to dMRI space using masks provided by the HCP-YA dataset to align the T1w and dMRI data. Macroscopic tissue response functions for white matter, gray matter, and cerebrospinal fluid were estimated from the dMRI data \cite{Dhollander2016, Dhollander2019, Tournier2004}. Multi-shell multi-tissue spherical deconvolution (MSMT-CSD) was applied to estimate FODs \cite{Jeurissen2014}, which were subsequently normalized to mitigate intensity differences in the signal \cite{Dhollander2021}.

The normalized FODs and the GMWM interface were then used to generate streamlines using probabilistic tractography, employing second-order integration over FODs (iFOD2) \cite{Tournier2010}. Ten million seeds were placed randomly throughout the GMWM interface for each subject. Streamlines that did not meet the ACT priors or exceeded a maximum length of 250 mm were rejected. The resulting streamlines were registered to MNI space using the nonlinear transform provided by the HCP-YA dataset. On average, this process produced 2.93 million streamlines per subject, with a standard deviation of 0.28 million.

\subsubsection{Ground truth connectome matrices and streamline node pair labels} \label{sec:2.2.2}
Structural connectome matrices and files containing per-streamline node pair labels were generated for each whole-brain tractogram using the tck2connectome command from MRtrix3 \cite{Tournier2019}. This command assigns a pair of gray matter regions (node pairs) to each streamline by calculating the distance from each streamline endpoint to the nearest parcellated gray matter region. For example, a streamline with endpoints closest to the left inferior parietal and right superior frontal regions will be labeled with the corresponding node indices (e.g., 7 and 76). The resulting node pair labels allow for the construction of a subject-specific connectivity matrix, where each entry in the matrix indicates the number of streamlines connecting a specific pair of regions. These matrices are symmetric, as the order of points in a streamline (forward or reverse) is irrelevant. For each subject, two sets of streamline node pair labels and connectivity matrices were generated, one using the 84-region and one using the 164-region parcellation provided by the HCP-YA dataset. 

\subsection{Connectome prediction method} \label{sec:2.3}
Our overall strategy for predicting connectomes from whole-brain tractograms was to classify individual streamlines based on the gray matter region pairs they connect. This was achieved by framing the task as a classification problem, where each streamline is treated as input, and the corresponding pair of gray matter nodes represents the ground truth class label. This section details the definition of streamline classes (~\ref{sec:2.3.1}) and the model architecture (~\ref{sec:2.3.2}) employed.

\subsubsection{Definition of streamline classes for prediction} \label{sec:2.3.1}
The streamline node pair labels, which included a pair of nodes for each streamline, were used to define streamline classes, such that each streamline received one class label. To determine the class labels, we used a systematic approach similar to lexicographical ordering \cite{Fishburn1974}, where pairs of regions are ordered by their indices to ensure that each unique pair is assigned a single class label. For any pair of connected regions, we assigned the same class label regardless of the order in which they appear (i.e., a streamline connecting region A to region B is treated the same as one connecting B to A). This approach ensures that each unique pair of regions has one corresponding class label. The number of possible classes varies with the parcellation scheme. The 84 and 164-region parcellations yield 3,571 and 13,631 possible classes (following \( n_{\text{classes}} = \frac{n_{\text{regions}} (n_{\text{regions}} + 1)}{2} + 1 \)). An additional “unknown” (label 0) class was included to account for any streamlines that tck2connectome was unable to assign to any parcellation node. This includes cases where streamlines terminate at the edge of the image, such as those that would otherwise be part of the corticospinal tract.

\subsubsection{Multi-task model architecture} \label{sec:2.3.2}
To classify individual streamlines into predefined classes, we adopted a state-of-the-art point-cloud-based method initially developed for tractography parcellation into anatomical tracts \cite{Xue2023-a}. Streamlines are represented as a set of 15 evenly spaced 3D coordinates along their trajectories 
\cite{Zhang2018-a, Zhang2020}. This representation enables the use of point-cloud neural networks, which are well-suited for processing unordered sets of spatial points. 

Our method builds upon the TractCloud model, which integrates both local and global streamline representations to enhance classification performance \cite{Xue2023-a}. The local representation encodes information from neighboring streamlines, capturing regional context, while the global representation provides information about the general brain pose. The original TractCloud architecture explored two widely used point-cloud neural networks, namely PointNet \cite{Qi-Charles2017} and Dynamic Graph Convolutional Neural Network (DGCNN) \cite{Wang2019}. PointNet processes each streamline’s points independently to extract point-wise features. To preserve anatomically important information about the spatial position of streamlines, the spatial transformation layer (T-Net) was removed from PointNet in this implementation \cite{Xue2023-b}. In contrast, DGCNN models interactions between points using a dynamic graph structure, capturing spatial relationships within each streamline. 

\begin{figure}[H]
    \centering
    \includegraphics[width=1\linewidth]{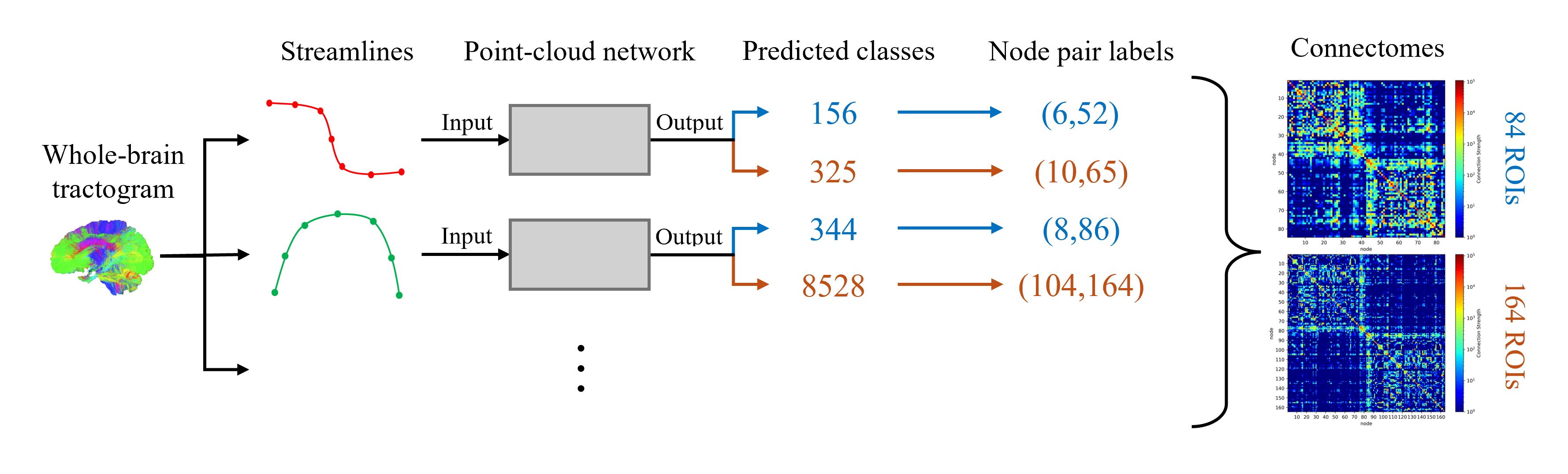}
    \caption{Diagram of the inference procedure of DeepMultiConnectome. One single streamline is used as input for the model, which predicts a class that is associated with the node pair it connects.}
    \label{fig:3}
\end{figure}

Besides adapting TractCloud to our problem using different tractography data and class labels, we employed a multi-task learning approach by adjusting the output layer to make multiple predictions simultaneously. Specifically, the model predicts two streamline classes, for both the 84 ROIs and 164 ROIs parcellation schemes, for each streamline in parallel. This was implemented by modifying the output layer to include separate fully connected layers for each parcellation scheme, both receiving input from the same learned streamline representation. This multi-task design, referred to as DeepMultiConnectome (Figure~\ref{fig:3}), leverages the shared streamline representation while enabling independent classification for each parcellation scheme.

Predictions of all the streamlines in a whole-brain tractogram are summarized in a connectome matrix. This figure shows the simultaneous prediction of two sets of node pairs according to two different parcellation schemes, using the multi-task learning approach.

\subsection{Implementation details}  \label{sec:2.4}
For model training, validation, and testing, we used a 70\%/10\%/20\% split of the 1000 subject dataset. The final optimized model was evaluated by predicting whole-brain connectomes for all 200 test subjects, using their complete sets of streamlines. Model training used 8 million streamlines from the training dataset, including 10,000 streamlines per training subject \cite{Xue2023-a, Xue2023-b}. The model was trained using cross-entropy loss and the Adam optimizer with a learning rate of 0.001 (selected from 0.01, 0.001, 0.0001) and a batch size of 1,024 (selected from 512, 1,024, and 2,048). Models were trained for 150 epochs, which took approximately 20 hours on a Jetstream2 Exouser instance with an NVIDIA A100 GPU \cite{Boerner2023, Hancock2021}.

\section{Results} \label{sec:3}
This section outlines our experiments aimed at optimizing model design (Section~\ref{sec:3.1}), evaluating the similarity between predicted and traditional connectomes (Section~\ref{sec:3.2}), and testing reproducibility using a test-retest dataset (Section~\ref{sec:3.3}). Additionally, we assess the practical utility of the predicted connectomes by evaluating their performance in a downstream age and cognitive performance prediction task (Section~\ref{sec:3.4}).

\subsection{Model design for computational efficiency} \label{sec:3.1}
Our overall goal is to predict multiple connectomes in an efficient manner, and therefore we performed two ablation-style experiments to optimize the DeepMultiConnectome model design using the 100-subject validation dataset. In these experiments, streamline classification performance was assessed using accuracy, macro F1, and computational efficiency (inference time per streamline). These metrics are widely used in tractography parcellation tasks \cite{Lam2018, Liu2019-af, Xue2023-a, Xue2023-b, Zhang2020}.

First, we focused on the task of predicting the connectome using the 84 ROI gray matter parcellation, and we compared two point-cloud backbone architectures, namely PointNet and DGCNN. We evaluated performance with and without the TractCloud local-global streamline embedding. A grid search was used to select the number of streamlines for the local-global representation \cite{Xue2023-a}, resulting in 50 local (from 10, 20, 50, 100) and 500 global streamlines (from 100, 300, 500, 1000). The results, summarized in Table~\ref{tab:1}, highlight trade-offs between model performance and computational efficiency. The DGCNN backbone, with a single streamline representation, achieved the highest accuracy (81.82\%) but had the lowest macro F1 score (47.6\%). In contrast, the PointNet backbone with the local-global streamline embedding attained the highest F1 score (51.01\%) but came with a substantial computational cost, requiring an inference time nearly ten times longer than the most efficient model. 

Given that our overall goal is to predict multiple connectomes in an efficient manner, we prioritized computational efficiency and selected the fastest backbone architecture (PointNet) for DeepMultiConnectome. The PointNet backbone with a single streamline representation offered a balanced trade-off, achieving a competitive accuracy (81.04\%) and the second-best F1 score (50.48\%), while maintaining the highest efficiency with a median inference time of 6.22~$\mu$s per streamline. All further experiments in the paper leverage this efficient PointNet backbone architecture.

\begin{table}[h]
\centering
\caption{Efficiency and performance of backbone architectures and streamline representations for predicting connectomes for the 84 ROI parcellation. Bold values indicate the most efficient architecture. The inference time per streamline is reported as median and interquartile range.}
\label{tab:1}
\begin{tabular}{|l|l|l|l|l|}
\hline
\makecell{Backbone \\ architecture} & \makecell{Streamline \\ representation} & \makecell{Accuracy} & \makecell{Macro F1} & \makecell{\textbf{Inference time} \\ \textbf{per streamline}} \\
\hline
\multirow{2}{*}{PointNet} 
  & Single               & 81.04\% & 50.48\% & \textbf{6.97 (4.97–10.70)~$\mu$s} \\
  \cline{2-5}
  & Local + Global       & 81.42\% & 51.01\% & 74.82 (72.25–77.94)~$\mu$s \\
\hline
\multirow{2}{*}{DGCNN} 
  & Single               & 81.82\% & 47.60\% & 14.81 (12.23–17.39)~$\mu$s \\
  \cline{2-5}
  & Local + Global       & 81.68\% & 48.83\% & 73.47 (69.06–76.83)~$\mu$s \\
\hline
\end{tabular}
\end{table}
 
Next, we assessed the efficiency and performance of single-task and multi-task models in predicting multiple connectomes from the 84 and 164 ROI parcellations. As shown in Table~\ref{tab:2}, the multi-task model achieved similar performance to the single-task models for both parcellation schemes while maintaining a low inference time per streamline. This shows that the multi-task model is highly efficient for predicting the two sets of class labels simultaneously. The rest of the experiments in this paper use the efficient design-optimized multi-task model, which is referred to as DeepMultiConnectome.

\begin{table}[h]
\centering
\caption{Efficiency and performance for single-task and multi-task models on 84 and 164 ROI parcellations. The inference time per streamline is reported as median and interquartile range. The total inference time per streamline refers to the time it takes to predict the node pairs for both parcellation schemes for a streamline.}
\label{tab:2}
\begin{tabular}{|l|l|l|l|l|l|}
\hline
\makecell{Model} & \makecell{Parcellation \\ scheme} & \makecell{Accuracy} & \makecell{Macro F1} & \makecell{Inference time \\ per streamline} & \makecell{\textbf{Total inference time} \\ \textbf{per streamline}} \\
\hline
\multirow{2}{*}{Single-Task} 
  & 84 ROI   & 81.04\% & 50.48\% & 6.97 (4.97–10.70)~$\mu$s & \multirow{2}{*}{16.48 (11.21–21.40)~$\mu$s} \\
  \cline{2-5}
  & 164 ROI  & 70.25\% & 35.03\% & 8.48 (6.08–12.05)~$\mu$s & \\
\hline
\multirow{2}{*}{Multi-Task} 
  & 84 ROI   & 81.23\% & 49.34\% & \multirow{2}{*}{7.51 (7.44–7.56)~$\mu$s} & \multirow{2}{*}{\textbf{7.51 (7.44–7.56)~$\mu$s}} \\
  \cline{2-4}
  & 164 ROI  & 70.15\% & 34.36\% & & \\
\hline
\end{tabular}
\end{table}

\subsection{Performance evaluation of connectome prediction} \label{sec:3.2}

Using the DeepMultiConnectome model, we predicted connectomes from whole-brain tractograms for the 200 unseen test subjects, each containing an average of 2.93 million streamlines. Connectome similarity between predicted and connectomes generated using the traditional pipeline was quantified using two complementary metrics, following the recommendations of \cite{Zalesky2024} : (1) Pearson's correlation, calculated between the upper triangular elements of the connectome matrices \cite{Osmanlioglu2020}, and (2) the log-Euclidean Riemannian metric (LERM), which measures the Riemannian distance between symmetric positive definite matrices \cite{Arsigny2006}. The intrasubject (within-subject) similarity was compared with intersubject (across-subject) similarity to determine if the intrasubject similarity was higher than the inherently high similarity of structural connectomes across subjects.

\begin{figure}[H]
    \centering
    \includegraphics[width=0.8\linewidth]{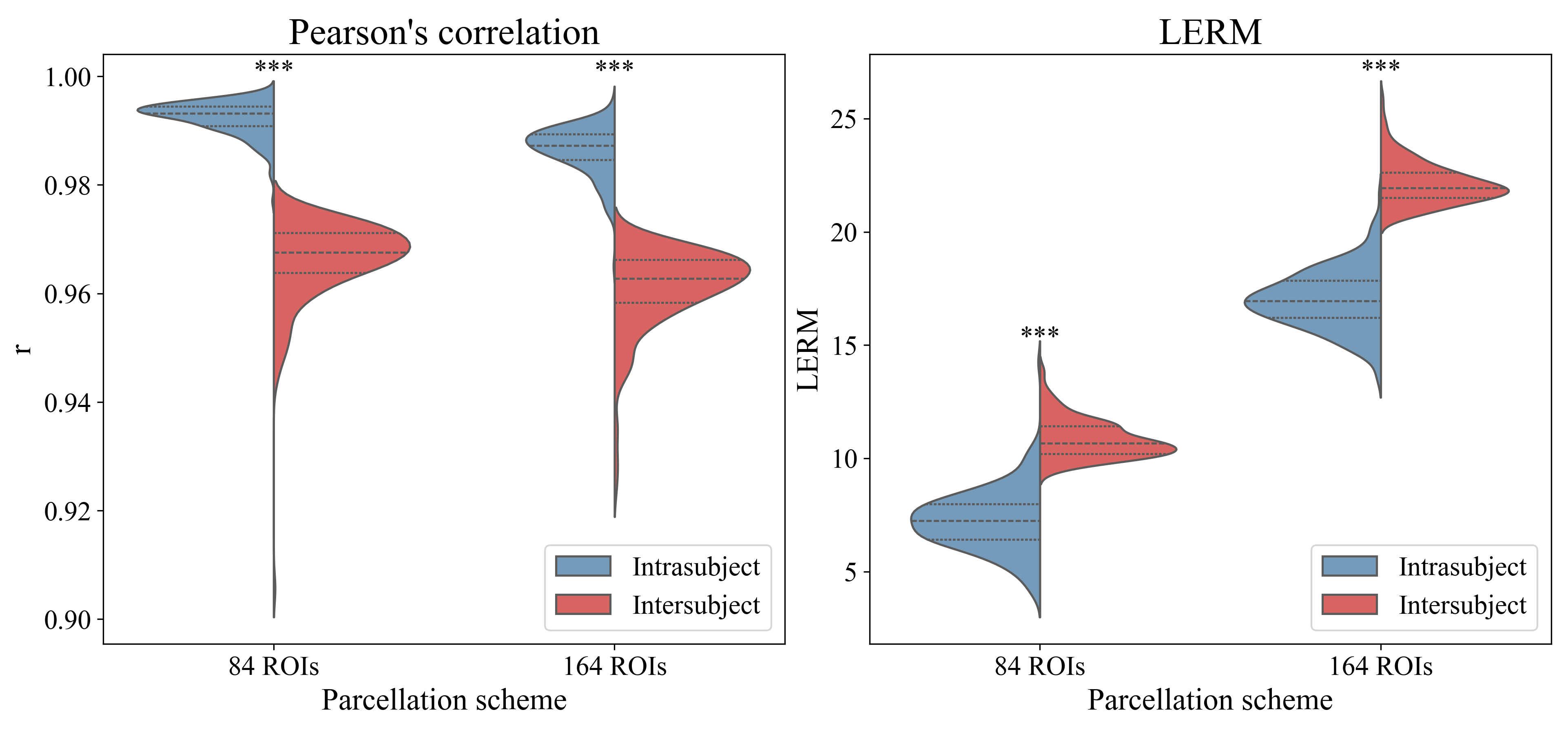}
    \caption{Similarity of predicted and traditional connectomes. Violin plots show the distributions of intrasubject (blue) and intersubject (red) correlation coefficients and log-Euclidean Riemannian metric (LERM) distances ($n=200$). Connectomes of both parcellation schemes are compared. *** denotes a statistically significant difference at \textit{p} $<$ 0.001 (Wilcoxon signed-rank test).}
    \label{fig:4}
\end{figure}

Inference required an average of 40.81 ± 5.06 seconds per tractogram. The predicted connectomes showed high similarity to their traditional counterparts, achieving a Pearson’s correlation of \textit{r }= 0.992 ± 0.003 and \textit{r }= 0.986 ± 0.004 for the 84 and 164 ROI parcellations, respectively. As expected, the intrasubject similarity was significantly higher than intersubject similarity across both measures and parcellation schemes (Figure~\ref{fig:4}). LERM distances were 7.25 ± 1.35 and 17.03 ± 1.29 for the two parcellations. Also, as expected, these intrasubject LERM distances were significantly lower than intersubject distances. This demonstrates that the predicted connectomes much more closely resembled their corresponding traditional connectomes than they resembled connectomes from other subjects (Figure~\ref{fig:4}).

\begin{figure}[H]
    \centering
    \includegraphics[width=1\linewidth]{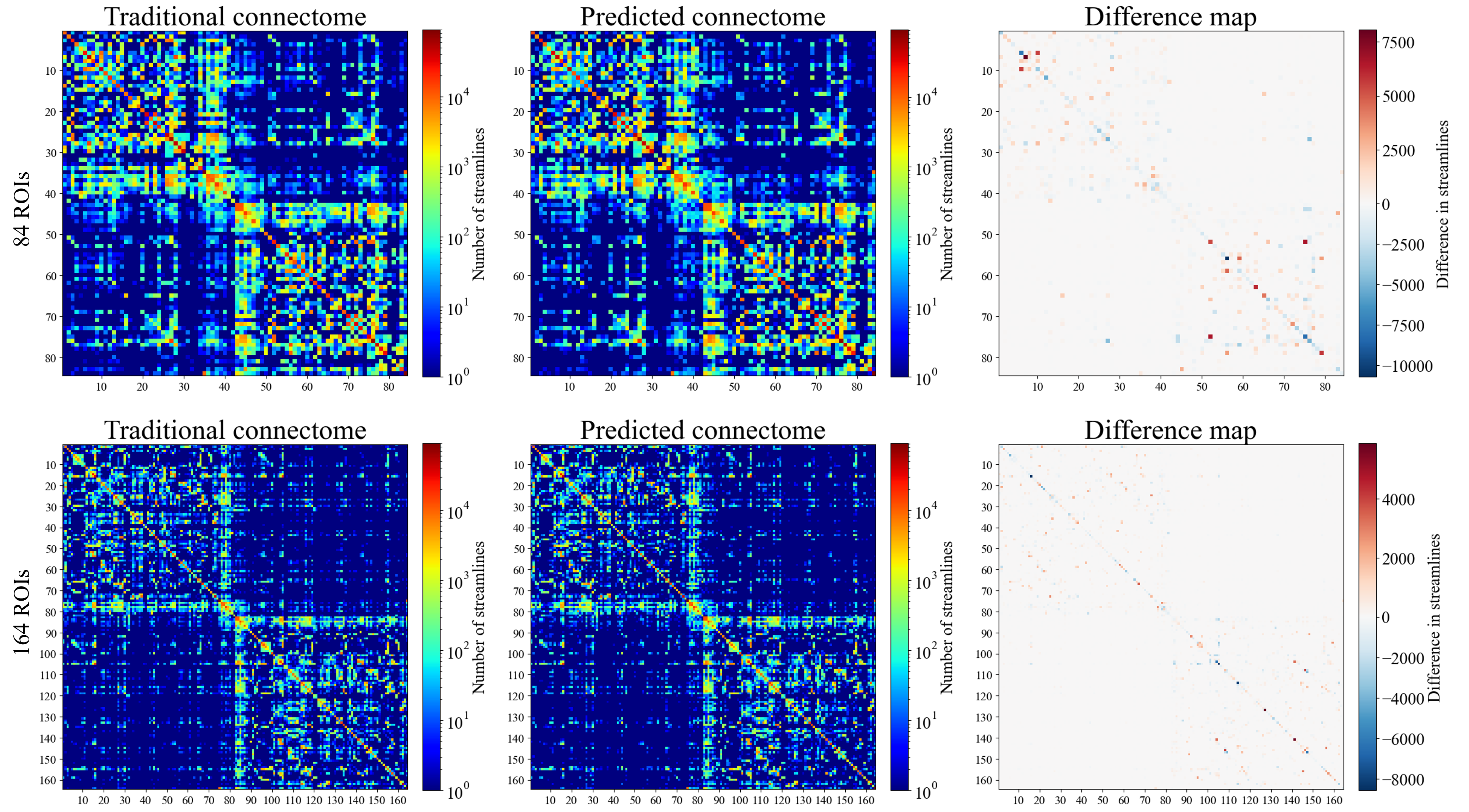}
    \caption{Traditional and DeepMultiConnectome predicted connectomes based on the 84 and 164 ROI parcellation schemes of an example test subject. The difference map was created by subtracting the predicted from the traditional connectome, and it shows the over- or underclassification of streamlines to each edge.}
    \label{fig:5}
\end{figure}

Visualizations of example structural connectomes, including a difference map highlighting discrepancies between predicted and ground truth connectomes, are provided in (Figure~\ref{fig:5}). Visually, predicted and ground truth connectomes appear highly similar, while the difference map reveals localized errors.
 
To evaluate the preservation of brain network organization in predicted connectomes, we assessed six widely used network measures. These measures are frequently employed in neuroimaging studies to characterize fundamental aspects of brain connectivity, including functional integration, segregation, and network resilience \cite{Rubinov2010}. Specifically, characteristic path length and global efficiency were analyzed as measures of integration; clustering coefficient, local efficiency, and modularity as measures of segregation; and assortativity as a measure of resilience. Table~\ref{tab:3} shows significant correlations between brain network measures computed from traditional and predicted connectomes. 

\begin{table}[h]
\centering
\caption{Pearson’s correlation and p-values between network measures computed from traditional and predicted connectomes for both parcellation schemes.}
\label{tab:3}
\begin{tabular}{|l|l|l|l|l|}
\hline
\multirow{2}{*}{\makecell{Network measure}} & \multicolumn{2}{|c|}{84 ROI} & \multicolumn{2}{|c|}{164 ROI} \\
\cline{2-5}
& \textit{r} & p-value & \textit{r} & p-value \\
\hline
Characteristic path length & 0.825 & $<$0.01 & 0.963 & $<$0.01 \\
Global efficiency           & 0.979 & $<$0.01 & 0.956 & $<$0.01 \\
Clustering coefficient      & 0.968 & $<$0.01 & 0.960 & $<$0.01 \\
Local efficiency            & 0.936 & $<$0.01 & 0.860 & $<$0.01 \\
Modularity                  & 0.963 & $<$0.01 & 0.918 & $<$0.01 \\
Assortativity               & 0.670 & $<$0.01 & 0.771 & $<$0.01 \\
\hline
\end{tabular}
\end{table}

\subsection{Test-retest reproducibility of predicted connectomes} \label{sec:3.3}
Reproducibility was assessed using the HCP-YA test-retest dataset, which includes 45 subjects scanned twice using an identical scan protocol, with a mean interval between scan sessions of approximately 140 days \cite{VanEssen2013}. Connectomes were generated for both scan sessions of each subject using both DeepMultiConnectome and the traditional method, and test-retest results were evaluated using Pearson’s correlation and LERM. 

The traditional and DeepMultiConnectome predicted connectomes showed similar test-retest reproducibility, with similar (statistically nondifferent) correlations and LERM values (Table~\ref{tab:4}). This was consistently observed for both 84 and 164 ROI parcellation schemes.

\begin{table}[h]
\centering
\caption{Reproducibility of connectomes generated with the traditional and DeepMultiConnectome methods. Evaluation of test-retest subject measures using Pearson’s correlation and log-Euclidean Riemannian metric (LERM) distance ($n=45$).}
\label{tab:4}
\begin{tabular}{|l|l|l|l|}
\hline
\makecell{Parcellation \\ scheme} & Connectome & \makecell{Pearson’s \\ correlation} & LERM \\
\hline
\multirow{2}{*}{84 ROI} 
  & Traditional           & 0.987 ± 0.007 & 8.897 ± 1.706 \\
  \cline{2-4}
  & DeepMultiConnectome   & 0.987 ± 0.008 & 8.858 ± 2.107 \\
\hline
\multirow{2}{*}{164 ROI} 
  & Traditional           & 0.981 ± 0.006 & 18.353 ± 1.479 \\
  \cline{2-4}
  & DeepMultiConnectome   & 0.980 ± 0.007 & 18.184 ± 1.613 \\
\hline
\end{tabular}
\end{table}

\subsection{Prediction of age and cognitive performance using connectomes} \label{sec:3.4}
Structural connectomes are often analyzed to predict individual non-imaging phenotypes related to cognition or brain age \cite{Baecker2021, Bernstein-Eliav2024, Shen2017, Soumya_Kumari2024}. In this experiment, we assessed the performance of the DeepMultiConnectome-generated connectomes in predicting age and human cognitive performance. We used cognitive performance measures from the NIH Toolbox, as described in Section~\ref{sec:2.1}. 

To model these predictions, we employed a deep network which has previously been successful in predicting non-imaging phenotypes from dMRI tractography data (\cite{Chen2023-b, He2022, Liu2023, Lo2024-b}). As input to the one-dimensional convolutional neural network (1D-CNN), we flattened the upper triangular parts of each connectome into a one-dimensional vector. The 1D-CNN architecture consisted of three convolutional layers with 64 filters and a kernel size of 5, followed by two fully connected layers with 512 and 128 neurons, respectively.

We evaluated the predictive performance of DeepMultiConnectome-generated connectomes against those generated using the traditional pipeline across both parcellation schemes. We used data from the 200 test subjects, as in Section~\ref{sec:3.2}, and trained and tested the 1D-CNN separately for each combination of connectome generation method and parcellation scheme using five-fold cross-validation. Predictive performance was evaluated using mean absolute error (MAE) for age prediction and Pearson's correlation coefficient (\textit{r})) for cognitive performance measures (Lo et al., 2025). The reported MAE and correlation values were computed by first calculating performance metrics separately for each of the five folds and then averaging these values across folds. 

As shown in Table~\ref{tab:5}, the predicted connectomes demonstrated predictive power comparable to their traditional counterparts. For the 84 ROI parcellation, the predicted connectomes achieved higher correlations in four out of seven measures, while for the 164 ROI parcellation, the predicted connectomes achieved higher correlations in two measures. Averaging these correlation scores (bottom row of Table~\ref{tab:5}) highlights the minimal performance differences between connectome generation methods. This suggests that the predicted connectomes effectively capture subject-specific information for age and cognitive function predictions, at a level comparable to traditionally generated connectomes.
Table~\ref{tab:5}: Prediction performance of traditional and predicted connectomes on age and various NIH toolbox measures used to evaluate cognition and language abilities. Predictive performance is measured in mean absolute error (MAE) for age and Pearson's correlation coefficient (\textit{r}) for the NIH Toolbox measures.

\newcolumntype{Y}{>{\centering\arraybackslash}X} 
\begin{table}[h]
\centering
\caption{Prediction performance of traditional and DeepMultiConnectome-based connectomes on age and various NIH Toolbox measures assessing cognition and language. Performance is measured as mean absolute error (MAE) for age and Pearson’s correlation coefficient (\textit{r}) for NIH Toolbox measures.}
\label{tab:5}
\begin{tabularx}{0.8\textwidth}{|l|Y|Y|Y|Y|}
\hline
\multirow{4}{*}{\makecell{Non-imaging \\ measure}} 
  & \multicolumn{4}{c|}{Parcellation scheme} \\
\cline{2-5}
  & \multicolumn{2}{c|}{84 ROI} & \multicolumn{2}{c|}{164 ROI} \\
\cline{2-5}
  & \makecell{Traditional} & \makecell{DeepMulti\\Connectome} & \makecell{Traditional} & \makecell{DeepMulti\\Connectome} \\
\hline
Age (MAE) & 3.47 ± 0.49 & 3.64 ± 0.70 & 4.04 ± 0.55 & 4.57 ± 0.76 \\
\hline
TPVT (\textit{r}) & 0.18 ± 0.17 & 0.19 ± 0.18 & 0.31 ± 0.19 & 0.28 ± 0.19 \\
\hline
TORRT (\textit{r}) & 0.21 ± 0.15 & 0.21 ± 0.16 & 0.20 ± 0.15 & 0.20 ± 0.13 \\
\hline
TFAT (\textit{r}) & 0.20 ± 0.07 & 0.21 ± 0.06 & 0.25 ± 0.11 & 0.24 ± 0.06 \\
\hline
TCST (\textit{r}) & 0.13 ± 0.10 & 0.17 ± 0.12 & 0.20 ± 0.14 & 0.18 ± 0.13 \\
\hline
TLST (\textit{r}) & 0.22 ± 0.16 & 0.24 ± 0.15 & 0.18 ± 0.15 & 0.28 ± 0.13 \\
\hline
TPIT (\textit{r}) & 0.21 ± 0.16 & 0.21 ± 0.13 & 0.16 ± 0.14 & 0.22 ± 0.21 \\
\hline
Average \textit{r} & 0.19 ± 0.14 & 0.20 ± 0.13 & 0.22 ± 0.15 & 0.23 ± 0.14 \\
\hline
\end{tabularx}
\end{table}

\section{Discussion and Conclusion}  \label{sec:4}
In this work, we introduced a deep learning approach to predict structural connectomes directly from tractography, bypassing gray matter parcellation. Our model, named DeepMultiConnectome, efficiently predicts structural connectomes across multiple parcellation schemes, enabling rapid, large-scale connectome generation while preserving subject-specific information critical for cognitive and neurological research. Key observations and implications from our experimental results are discussed below.

Our model demonstrates substantial improvements in computational efficiency. By leveraging previous work that optimized streamline classification speed and optimizing the model design in this study, we achieved an average prediction time of 40 seconds per whole-brain tractogram with approximately 3 million streamlines. Additionally, the multi-task model allows simultaneous predictions across multiple parcellation schemes, reducing the need for separate atlas-specific preprocessing steps. In contrast, traditional techniques like FreeSurfer \cite{Fischl2012} require hours per subject to match streamline endpoints to ROIs. While methods like FastSurfer \cite{Henschel2020} significantly accelerate gray matter parcellation, its current implementation is limited to one atlas. Our approach thus offers a scalable alternative, particularly advantageous for large neuroimaging studies.

The predicted structural connectomes using DeepMultiConnectome show high similarity with traditional connectomes derived from FreeSurfer parcellations, measured as the correlation and distance between connectomes and the correlation of derived network measures. The predicted structural connectomes also exhibit reproducibility comparable to the traditional method when evaluated on test-retest data and produce downstream predictions on non-imaging measures at a level similar to traditional connectomes. It is notable that the connectomes achieve such promising consistency and downstream performance despite relatively modest streamline classification accuracy and macro F1 scores. One possible explanation is that misclassifications occur in regions with limited impact on overall network structure or downstream predictions. This interpretation is supported by the difference maps, which reveal few edges with substantial classification errors. Consequently, the overall connectivity patterns remain largely intact. Such localized inaccuracies could account for the high accuracy yet low macro F1 scores, as errors are concentrated in a limited subset of classes. Alternatively, our evaluation methods may not be sufficiently sensitive, potentially overestimating the true quality of the predictions.

Recent deep-learning approaches have also explored alternative strategies for structural connectome prediction. \cite{Liu2025} proposed generating personalized structural connectomes in the absence of MRI data, using only non-imaging phenotype data such as age, sex, and cognitive measures. The authors demonstrated the utility of their approach in data augmentation to reduce predictive errors. Alternatively, \cite{Sarwar2020} proposed a strategy to synthesize structural connectomes by combining information from multiple small-scale local connectivity matrices that could be predicted from dMRI data without whole-brain tractography. While these deep learning methods differ from our approach, which directly leverages tractography data, they collectively broaden the landscape of connectome generation strategies.

We note some limitations of this study and directions for future research. A key challenge is the high anatomical variability among individuals, particularly in complex gyral patterns. Such variability may cause streamlines with similar shapes to end in adjacent, yet distinct regions, leading to misclassifications. To address this issue, future work could explore alternative modeling strategies, such as independently predicting each endpoint’s node label. This approach would reduce the total number of classes, potentially improving classification accuracy. Additionally, we did not investigate the model’s ability to generalize to a broader population across the lifespan. Future work could assess generalization performance and potentially incorporate more varied training data, such as different acquisitions and tractography methods. Furthermore, incorporating other more complex gray matter parcellation schemes and evaluating results with alternative network weighting strategies (e.g., fractional anisotropy or SIFT2) may further refine and broaden the applicability of the proposed DeepMultiConnectome model. 

In conclusion, this study presents a time-efficient deep learning method for predicting structural connectomes directly from tractography, bypassing the gray matter parcellation process. Our method predicts connectomes with high similarity to traditionally generated connectomes while strongly reducing computation times. This work can provide a scalable, rapid model for producing subject-specific connectomes across diverse gray matter parcellation schemes.
 
\needspace{4\baselineskip}
\section*{Acknowledgments}
This work used \textbf{Jetstream2} through allocation MED230035 from the Advanced Cyberinfrastructure Coordination Ecosystem: Services \& Support (ACCESS) program, supported by National Science Foundation grants \#2138259, \#2138286, \#2138307, \#2137603, and \#2138296.

\needspace{1\baselineskip}
\section*{Conflict of Interest}
The authors declare no conflict of interest.

\needspace{4\baselineskip}
\section*{Data Availability}
The HCP-YA dataset used in this project can be downloaded through the ConnectomeDB (\url{https://db.humanconnectome.org}). All code developed for our experiments will be publicly available at: \url{https://github.com/SlicerDMRI/DeepMultiConnectome}.

\nocite{*}

\bibliographystyle{apalike}
\bibliography{references}

\begin{thebibliography}{}

\bibitem[Arsigny et~al., 2006]{Arsigny2006}
Arsigny, V., Fillard, P., Pennec, X., and Ayache, N. (2006).
\newblock Log-euclidean metrics for fast and simple calculus on diffusion
  tensors.
\newblock {\em Magn. Reson. Med.}, 56(2):411--421.

\bibitem[Arslan et~al., 2018]{Arslan2018}
Arslan, S., Ktena, S.~I., Makropoulos, A., Robinson, E.~C., Rueckert, D., and
  Parisot, S. (2018).
\newblock Human brain mapping: A systematic comparison of parcellation methods
  for the human cerebral cortex.
\newblock {\em Neuroimage}, 170:5--30.

\bibitem[Astolfi et~al., 2020]{Astolfi2020}
Astolfi, P., Verhagen, R., Petit, L., Olivetti, E., Masci, J., Boscaini, D.,
  and Avesani, P. (2020).
\newblock Tractogram filtering of anatomically non-plausible fibers with
  geometric deep learning.
\newblock In {\em Medical Image Computing and Computer Assisted Intervention
  – MICCAI 2020}, Lecture notes in computer science, pages 291--301. Springer
  International Publishing, Cham.

\bibitem[Baecker et~al., 2021]{Baecker2021}
Baecker, L., Garcia-Dias, R., Vieira, S., Scarpazza, C., and Mechelli, A.
  (2021).
\newblock Machine learning for brain age prediction: Introduction to methods
  and clinical applications.
\newblock {\em EBioMedicine}, 72(103600):103600.

\bibitem[Bassett and Bullmore, 2009]{Bassett2009}
Bassett, D.~S. and Bullmore, E.~T. (2009).
\newblock Human brain networks in health and disease.
\newblock {\em Curr. Opin. Neurol.}, 22(4):340--347.

\bibitem[Bernstein-Eliav and Tavor, 2024]{Bernstein-Eliav2024}
Bernstein-Eliav, M. and Tavor, I. (2024).
\newblock The prediction of brain activity from connectivity: Advances and
  applications.
\newblock {\em Neuroscientist}, 30(3):367--377.

\bibitem[Boerner et~al., 2023]{Boerner2023}
Boerner, T.~J., Deems, S., Furlani, T.~R., Knuth, S.~L., and Towns, J. (2023).
\newblock {ACCESS}: Advancing innovation: {NSF’s} advanced
  cyberinfrastructure coordination ecosystem: Services \& support.
\newblock In {\em Practice and Experience in Advanced Research Computing}, New
  York, NY, USA. ACM.

\bibitem[Cao et~al., 2016]{Cao2016}
Cao, M., Huang, H., Peng, Y., Dong, Q., and He, Y. (2016).
\newblock Toward developmental connectomics of the human brain.
\newblock {\em Front. Neuroanat.}, 10:25.

\bibitem[Chen et~al., 2023a]{Chen2023-b}
Chen, Y., Zhang, C., Xue, T., Song, Y., Makris, N., Rathi, Y., Cai, W., Zhang,
  F., and O'Donnell, L.~J. (2023a).
\newblock Deep fiber clustering: Anatomically informed fiber clustering with
  self-supervised deep learning for fast and effective tractography
  parcellation.
\newblock {\em Neuroimage}, 273:120086.

\bibitem[Chen et~al., 2023b]{Chen2023-a}
Chen, Y., Zhang, F., Zekelman, L.~R., Xue, T., Zhang, C., Song, Y., Makris, N.,
  Rathi, Y., Cai, W., and O'Donnell, L.~J. (2023b).
\newblock Tractgraphcnn: Anatomically informed graph {CNN} for classification
  using diffusion {MRI} tractography.
\newblock In {\em 2023 IEEE 20th International Symposium on Biomedical Imaging
  (ISBI)}, pages 1--5. IEEE.

\bibitem[Chen et~al., 2022]{Chen2022}
Chen, Y., Zhang, F., Zhang, C., Xue, T., Zekelman, L.~R., He, J., Song, Y.,
  Makris, N., Rathi, Y., Golby, A.~J., Cai, W., and O'Donnell, L.~J. (2022).
\newblock White matter tracts are point clouds: Neuropsychological score
  prediction and critical region localization via geometric deep learning.
\newblock In {\em Lecture Notes in Computer Science}, Lecture notes in computer
  science, pages 174--184. Springer Nature Switzerland, Cham.

\bibitem[Desikan et~al., 2006]{Desikan2006}
Desikan, R.~S., Ségonne, F., Fischl, B., Quinn, B.~T., Dickerson, B.~C.,
  Blacker, D., Buckner, R.~L., Dale, A.~M., Maguire, R.~P., Hyman, B.~T.,
  Albert, M.~S., and Killiany, R.~J. (2006).
\newblock An automated labeling system for subdividing the human cerebral
  cortex on {MRI} scans into gyral based regions of interest.
\newblock {\em Neuroimage}, 31(3):968--980.

\bibitem[Destrieux et~al., 2010]{Destrieux2010}
Destrieux, C., Fischl, B., Dale, A., and Halgren, E. (2010).
\newblock Automatic parcellation of human cortical gyri and sulci using
  standard anatomical nomenclature.
\newblock {\em Neuroimage}, 53(1):1--15.

\bibitem[Dhollander et~al., 2019]{Dhollander2019}
Dhollander, T., Mito, R., Raffelt, D., and Connelly, A. (2019).
\newblock Improved white matter response function estimation for 3-tissue
  constrained spherical deconvolution.
\newblock In {\em Proc. Intl. Soc. Mag. Reson. Med}, volume 555.
  archive.ismrm.org.

\bibitem[Dhollander et~al., 2016]{Dhollander2016}
Dhollander, T., Raffelt, D., and Connelly, A. (2016).
\newblock Unsupervised 3-tissue response function estimation from single-shell
  or multi-shell diffusion {MR} data without a co-registered {T1} image.

\bibitem[Dhollander et~al., 2021]{Dhollander2021}
Dhollander, T., Tabbara, R., Rosnarho-Tornstrand, J., Tournier, J.-D., Raffelt,
  D., and Connelly, A. (2021).
\newblock Multi-tissue log-domain intensity and inhomogeneity normalisation for
  quantitative apparent fibre density.
\newblock In {\em Proc. ISMRM}, volume~29, page 2472. archive.ismrm.org.

\bibitem[Fischl, 2012]{Fischl2012}
Fischl, B. (2012).
\newblock {FreeSurfer}.
\newblock {\em Neuroimage}, 62(2):774--781.

\bibitem[Fishburn, 1974]{Fishburn1974}
Fishburn, P.~C. (1974).
\newblock Exceptional paper—lexicographic orders, utilities and decision
  rules: A survey.
\newblock {\em Manage. Sci.}, 20(11):1442--1471.

\bibitem[Gershon et~al., 2013]{Gershon2013}
Gershon, R.~C., Slotkin, J., Manly, J.~J., Blitz, D.~L., Beaumont, J.~L.,
  Schnipke, D., Wallner-Allen, K., Golinkoff, R.~M., Gleason, J.~B.,
  Hirsh-Pasek, K., Adams, M.~J., and Weintraub, S. (2013).
\newblock {IV}. {NIH} toolbox cognition battery ({CB}): measuring language
  (vocabulary comprehension and reading decoding): Nih toolbox cognition
  battery (cb).
\newblock {\em Monogr. Soc. Res. Child Dev.}, 78(4):49--69.

\bibitem[Glasser et~al., 2013]{Glasser2013}
Glasser, M.~F., Sotiropoulos, S.~N., Wilson, J.~A., Coalson, T.~S., Fischl, B.,
  Andersson, J.~L., Xu, J., Jbabdi, S., Webster, M., Polimeni, J.~R.,
  Van~Essen, D.~C., Jenkinson, M., and {WU-Minn HCP Consortium} (2013).
\newblock The minimal preprocessing pipelines for the human connectome project.
\newblock {\em Neuroimage}, 80:105--124.

\bibitem[Gruber et~al., 2023]{Gruber2023}
Gruber, M., Mauritz, M., Meinert, S., Grotegerd, D., de~Lange, S.~C., Grumbach,
  P., Goltermann, J., Winter, N.~R., Waltemate, L., Lemke, H., Thiel, K.,
  Winter, A., Breuer, F., Borgers, T., Enneking, V., Klug, M., Brosch, K.,
  Meller, T., Pfarr, J.-K., Ringwald, K.~G., Stein, F., Opel, N., Redlich, R.,
  Hahn, T., Leehr, E.~J., Bauer, J., Nenadić, I., Kircher, T., van~den Heuvel,
  M.~P., Dannlowski, U., and Repple, J. (2023).
\newblock Cognitive performance and brain structural connectome alterations in
  major depressive disorder.
\newblock {\em Psychol. Med.}, 53(14):6611--6622.

\bibitem[Hancock et~al., 2021]{Hancock2021}
Hancock, D.~Y., Fischer, J., Lowe, J.~M., Snapp-Childs, W., Pierce, M., Marru,
  S., Coulter, J.~E., Vaughn, M., Beck, B., Merchant, N., Skidmore, E., and
  Jacobs, G. (2021).
\newblock {Jetstream2}: Accelerating cloud computing via jetstream.
\newblock In {\em Practice and Experience in Advanced Research Computing}, New
  York, NY, USA. ACM.

\bibitem[He et~al., 2022]{He2022}
He, H., Zhang, F., Pieper, S., Makris, N., Rathi, Y., Wells, W., and O'Donnell,
  L.~J. (2022).
\newblock Model and predict age and sex in healthy subjects using brain white
  matter features: A deep learning approach.
\newblock In {\em 2022 IEEE 19th International Symposium on Biomedical Imaging
  (ISBI)}, pages 1--5. IEEE.

\bibitem[Henschel et~al., 2020]{Henschel2020}
Henschel, L., Conjeti, S., Estrada, S., Diers, K., Fischl, B., and Reuter, M.
  (2020).
\newblock {FastSurfer} - a fast and accurate deep learning based neuroimaging
  pipeline.
\newblock {\em Neuroimage}, 219(117012):117012.

\bibitem[Jeurissen et~al., 2014]{Jeurissen2014}
Jeurissen, B., Tournier, J.-D., Dhollander, T., Connelly, A., and Sijbers, J.
  (2014).
\newblock Multi-tissue constrained spherical deconvolution for improved
  analysis of multi-shell diffusion {MRI} data.
\newblock {\em Neuroimage}, 103:411--426.

\bibitem[Karimi et~al., 2024]{Karimi2024}
Karimi, D., Calixto, C., Snoussi, H., Cortes-Albornoz, M.~C., Velasco-Annis,
  C., Rollins, C., Jaimes, C., Gholipour, A., and Warfield, S.~K. (2024).
\newblock Detailed delineation of the fetal brain in diffusion {MRI} via
  multi-task learning.
\newblock {\em bioRxivorg}.

\bibitem[Lam et~al., 2018]{Lam2018}
Lam, N., Belhomme, P.~D., Ferrall, G., Patterson, J., Styner, B., and Prieto,
  M. (2018).
\newblock {TRAFIC}: Fiber tract classification using deep learning.
\newblock {\em Proc. SPIE Int. Soc. Opt. Eng}.

\bibitem[Liu et~al., 2019]{Liu2019-af}
Liu, F., Feng, J., Chen, G., Wu, Y., Hong, Y., Yap, P.-T., and Shen, D. (2019).
\newblock {DeepBundle}: Fiber bundle parcellation with graph convolution neural
  networks.
\newblock {\em arXiv [eess.IV]}.

\bibitem[Liu et~al., 2023]{Liu2023}
Liu, W., Chen, Y., Ye, C., Makris, N., Rathi, Y., Cai, W., Zhang, F., and
  O'Donnell, L.~J. (2023).
\newblock Fiber tract shape measures inform prediction of non-imaging
  phenotypes.
\newblock {\em arXiv [cs.CV]}.

\bibitem[Liu et~al., 2025]{Liu2025}
Liu, Y., Seguin, C., Mansour~L., S., Tian, Y., Di~Biase, M.~A., and Zalesky, A.
  (2025).
\newblock Deep generation of personalized connectomes based on individual
  attributes.
\newblock {\em bioRxiv}.

\bibitem[Lo et~al., 2024a]{Lo2024-a}
Lo, Y., Chen, Y., Liu, D., Legarreta, J.~H., Zekelman, L., Zhang, F., Rushmore,
  J., Rathi, Y., Makris, N., Golby, A.~J., Cai, W., and O'Donnell, L.~J.
  (2024a).
\newblock {TractShapeNet}: Efficient multi-shape learning with {3D}
  tractography point clouds.
\newblock {\em arXiv [cs.CV]}.

\bibitem[Lo et~al., 2025]{Lo2025}
Lo, Y., Chen, Y., Liu, D., Liu, W., Zekelman, L., Rushmore, J., Zhang, F.,
  Rathi, Y., Makris, N., Golby, A.~J., Cai, W., and O'Donnell, L.~J. (2025).
\newblock The shape of the brain's connections is predictive of cognitive
  performance: An explainable machine learning study.
\newblock {\em Hum. Brain Mapp.}, 46(5):e70166.

\bibitem[Lo et~al., 2024b]{Lo2024-b}
Lo, Y., Chen, Y., Liu, D., Liu, W., Zekelman, L., Zhang, F., Rathi, Y., Makris,
  N., Golby, A.~J., Cai, W., et~al. (2024b).
\newblock Cross-domain fiber cluster shape analysis for language performance
  cognitive score prediction.
\newblock In {\em International Workshop on Computational Diffusion MRI}, pages
  84--94. Springer.

\bibitem[Osmanlıoğlu et~al., 2020]{Osmanlioglu2020}
Osmanlıoğlu, Y., Alappatt, J.~A., Parker, D., and Verma, R. (2020).
\newblock Connectomic consistency: a systematic stability analysis of
  structural and functional connectivity.
\newblock {\em J. Neural Eng.}, 17(4):045004.

\bibitem[Qi~Charles et~al., 2017]{Qi-Charles2017}
Qi~Charles, R., Su, H., Kaichun, M., and Guibas, L.~J. (2017).
\newblock {PointNet}: Deep learning on point sets for {3D} classification and
  segmentation.
\newblock In {\em 2017 IEEE Conference on Computer Vision and Pattern
  Recognition (CVPR)}, pages 77--85. IEEE.

\bibitem[Rubinov and Sporns, 2010]{Rubinov2010}
Rubinov, M. and Sporns, O. (2010).
\newblock Complex network measures of brain connectivity: uses and
  interpretations.
\newblock {\em Neuroimage}, 52(3):1059--1069.

\bibitem[Sarwar et~al., 2020]{Sarwar2020}
Sarwar, T., Seguin, C., Ramamohanarao, K., and Zalesky, A. (2020).
\newblock Towards deep learning for connectome mapping: A block decomposition
  framework.
\newblock {\em Neuroimage}, 212(116654):116654.

\bibitem[Shen et~al., 2017]{Shen2017}
Shen, X., Finn, E.~S., Scheinost, D., Rosenberg, M.~D., Chun, M.~M.,
  Papademetris, X., and Constable, R.~T. (2017).
\newblock Using connectome-based predictive modeling to predict individual
  behavior from brain connectivity.
\newblock {\em Nat. Protoc.}, 12(3):506--518.

\bibitem[Smith et~al., 2012]{Smith2012}
Smith, R.~E., Tournier, J.-D., Calamante, F., and Connelly, A. (2012).
\newblock Anatomically-constrained tractography: improved diffusion {MRI}
  streamlines tractography through effective use of anatomical information.
\newblock {\em Neuroimage}, 62(3):1924--1938.

\bibitem[Sotiropoulos et~al., 2013]{Sotiropoulos2013}
Sotiropoulos, S.~N., Jbabdi, S., Xu, J., Andersson, J.~L., Moeller, S.,
  Auerbach, E.~J., Glasser, M.~F., Hernandez, M., Sapiro, G., Jenkinson, M.,
  Feinberg, D.~A., Yacoub, E., Lenglet, C., Van~Essen, D.~C., Ugurbil, K.,
  Behrens, T. E.~J., and {WU-Minn HCP Consortium} (2013).
\newblock Advances in diffusion {MRI} acquisition and processing in the human
  connectome project.
\newblock {\em Neuroimage}, 80:125--143.

\bibitem[Soumya~Kumari and Sundarrajan, 2024]{Soumya_Kumari2024}
Soumya~Kumari, L.~K. and Sundarrajan, R. (2024).
\newblock A review on brain age prediction models.
\newblock {\em Brain Res.}, 1823(148668):148668.

\bibitem[Sporns et~al., 2005]{Sporns2005}
Sporns, O., Tononi, G., and Kötter, R. (2005).
\newblock The human connectome: A structural description of the human brain.
\newblock {\em PLoS Comput. Biol.}, 1(4):e42.

\bibitem[Toga et~al., 2012]{Toga2012}
Toga, A.~W., Clark, K.~A., Thompson, P.~M., Shattuck, D.~W., and Van~Horn,
  J.~D. (2012).
\newblock Mapping the human connectome.
\newblock {\em Neurosurgery}, 71(1):1--5.

\bibitem[Tournier et~al., 2010]{Tournier2010}
Tournier, J.~D., Calamante, F., Connelly, A., and {Others} (2010).
\newblock Improved probabilistic streamlines tractography by {2nd} order
  integration over fibre orientation distributions.
\newblock In {\em Proceedings of the international society for magnetic
  resonance in medicine}, volume 1670. archive.ismrm.org.

\bibitem[Tournier et~al., 2004]{Tournier2004}
Tournier, J.-D., Calamante, F., Gadian, D.~G., and Connelly, A. (2004).
\newblock Direct estimation of the fiber orientation density function from
  diffusion-weighted {MRI} data using spherical deconvolution.
\newblock {\em Neuroimage}, 23(3):1176--1185.

\bibitem[Tournier et~al., 2019]{Tournier2019}
Tournier, J.-D., Smith, R., Raffelt, D., Tabbara, R., Dhollander, T., Pietsch,
  M., Christiaens, D., Jeurissen, B., Yeh, C.-H., and Connelly, A. (2019).
\newblock {MRtrix3}: A fast, flexible and open software framework for medical
  image processing and visualisation.
\newblock {\em Neuroimage}.

\bibitem[Van~Essen et~al., 2013]{VanEssen2013}
Van~Essen, D.~C., Smith, S.~M., Barch, D.~M., Behrens, T. E.~J., Yacoub, E.,
  Ugurbil, K., and {WU-Minn HCP Consortium} (2013).
\newblock The {WU}-minn human connectome project: an overview.
\newblock {\em Neuroimage}, 80:62--79.

\bibitem[Wainberg et~al., 2024]{Wainberg2024}
Wainberg, M., Forde, N.~J., Mansour, S., Kerrebijn, I., Medland, S.~E., Hawco,
  C., and Tripathy, S.~J. (2024).
\newblock Genetic architecture of the structural connectome.
\newblock {\em Nat. Commun.}, 15(1):1962.

\bibitem[Wang et~al., 2019]{Wang2019}
Wang, Y., Sun, Y., Liu, Z., Sarma, S.~E., Bronstein, M.~M., and Solomon, J.~M.
  (2019).
\newblock Dynamic graph {CNN} for learning on point clouds.
\newblock {\em ACM Trans. Graph.}, 38(5):1--12.

\bibitem[Wasserthal et~al., 2018]{Wasserthal2018}
Wasserthal, J., Neher, P., and Maier-Hein, K.~H. (2018).
\newblock {TractSeg} - fast and accurate white matter tract segmentation.
\newblock {\em Neuroimage}, 183:239--253.

\bibitem[Weintraub et~al., 2013]{Weintraub2013}
Weintraub, S., Dikmen, S.~S., Heaton, R.~K., Tulsky, D.~S., Zelazo, P.~D.,
  Bauer, P.~J., Carlozzi, N.~E., Slotkin, J., Blitz, D., Wallner-Allen, K.,
  Fox, N.~A., Beaumont, J.~L., Mungas, D., Nowinski, C.~J., Richler, J.,
  Deocampo, J.~A., Anderson, J.~E., Manly, J.~J., Borosh, B., Havlik, R.,
  Conway, K., Edwards, E., Freund, L., King, J.~W., Moy, C., Witt, E., and
  Gershon, R.~C. (2013).
\newblock Cognition assessment using the {NIH} toolbox.
\newblock {\em Neurology}, 80(11 Suppl 3):S54.

\bibitem[Xue et~al., 2023a]{Xue2023-a}
Xue, T., Chen, Y., Zhang, C., Golby, A.~J., Makris, N., Rathi, Y., Cai, W.,
  Zhang, F., and O'Donnell, L.~J. (2023a).
\newblock {TractCloud}: Registration-free tractography parcellation with a
  novel local-global streamline point cloud representation.
\newblock In {\em Lecture Notes in Computer Science}, Lecture notes in computer
  science, pages 409--419. Springer Nature Switzerland, Cham.

\bibitem[Xue et~al., 2023b]{Xue2023-b}
Xue, T., Zhang, F., Zhang, C., Chen, Y., Song, Y., Golby, A.~J., Makris, N.,
  Rathi, Y., Cai, W., and O'Donnell, L.~J. (2023b).
\newblock Superficial white matter analysis: An efficient point-cloud-based
  deep learning framework with supervised contrastive learning for consistent
  tractography parcellation across populations and {dMRI} acquisitions.
\newblock {\em Med. Image Anal.}, 85:102759.

\bibitem[Zalesky et~al., 2024]{Zalesky2024}
Zalesky, A., Sarwar, T., Tian, Y., Liu, Y., Yeo, B. T.~T., and Ramamohanarao,
  K. (2024).
\newblock Predicting an individual's functional connectivity from their
  structural connectome: Evaluation of evidence, recommendations, and future
  prospects.
\newblock {\em Netw. Neurosci.}, 8(4):1291--1309.

\bibitem[Zhang et~al., 2020]{Zhang2020}
Zhang, F., Cetin~Karayumak, S., Hoffmann, N., Rathi, Y., Golby, A.~J., and
  O'Donnell, L.~J. (2020).
\newblock Deep white matter analysis ({DeepWMA}): Fast and consistent
  tractography segmentation.
\newblock {\em Med. Image Anal.}, 65(101761):101761.

\bibitem[Zhang et~al., 2018]{Zhang2018-a}
Zhang, F., Wu, Y., Norton, I., Rigolo, L., Rathi, Y., Makris, N., and
  O'Donnell, L.~J. (2018).
\newblock An anatomically curated fiber clustering white matter atlas for
  consistent white matter tract parcellation across the lifespan.
\newblock {\em Neuroimage}, 179:429--447.

\bibitem[Zhang and Yang, 2018]{Zhang2018-b}
Zhang, Y. and Yang, Q. (2018).
\newblock An overview of multi-task learning.
\newblock {\em Natl. Sci. Rev.}, 5(1):30--43.

\end{thebibliography}

\end{document}